\documentclass[a4paper,11pt]{article}
\usepackage{pos}
\usepackage{graphicx} 
\usepackage{subcaption}
\usepackage{wrapfig}

\title{Searching for TeV emission from LHAASO\,J0341+5258 with VERITAS and HAWC}
 \ShortTitle{VERITAS and HAWC study of LHAASO\,J0341+5258}

\author*[a]{P. Bangale}
\author[]{for the VERITAS Collaboration}
\author[b]{X. Wang}
\onbehalf{for the HAWC Collaboration}

\affiliation[a]{Department of Physics and Astronomy and the Bartol Research Institute, \\ University of Delaware, Newark, DE 19716, USA} 
\affiliation[b]{Department of Physics, \\ Michigan Technological University, Houghton, MI 49931, USA}

\emailAdd{pbangale@udel.edu}

\abstract{Galactic PeVatrons are astrophysical sources accelerating particles up to a few PeV ($\sim$10$^{15}$\,eV) energies. The primary signature of 100 TeV $\gamma$ rays may come from PeV protons or multi-hundred TeV (not PeV) electrons.~The search for PeVatrons has been one of the key science topics for VERITAS and HAWC. In 2021, LHAASO detected 14 steady $\gamma$-ray sources with photon energies above 100\,TeV, up to 1.4\,PeV. This provides a clear list of PeVatron candidates for further study with VERITAS and HAWC. Most of these sources contain possible source associations, such as supernova remnants, pulsar wind nebulae, and stellar clusters. However, two sources: LHAASO\,J2108+5157 and LHAASO\,J0341+5258, do not have any such counterparts. Therefore, multiwavelength observations are required to identify the objects responsible for the UHE $\gamma$ rays, to understand the source morphology and association, and to shed light on the emission processes. Here, we will present the status of VERITAS/HAWC observations and results for the LHAASO PeVatron candidate J0341+5258, and also discuss the VERITAS PeVatron search in general.}

\ConferenceLogo{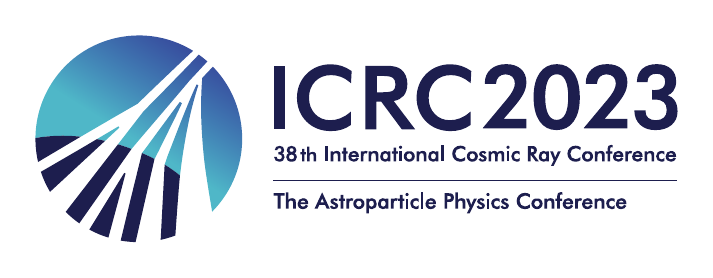}

\FullConference{%
38th International Cosmic Ray Conference (ICRC2023)\\
  26 July - 3 August, 2023\\
  Nagoya, Japan}

\begin{document}
\maketitle

\section{Introduction} \label{sec:intro}

The study of PeVatrons, particularly identifying hadronic PeVatrons, is a crucial step towards understanding the longstanding problem of cosmic ray origin.~$\gamma$-ray observations in the ultra-high energy (UHE, E$>$100\,TeV) band are essential for PeVatron searches.~The current generation of imaging atmospheric Cherenkov telescopes (IACTs) are most sensitive to $\gamma$-ray photons around TeV energies. To study $\gamma$ rays above $\sim$10\,TeV requires a large effective area, and a long exposure, which is limited by the moderate ﬁeld of view (FoV) and the duty cycle of current IACTs. Extensive air shower arrays, such as Tibet-As$\gamma$, HAWC and LHAASO, have a wide FoV and high-duty cycle, which is useful in detecting $\gamma$-ray sources beyond energies of tens of TeV, and up into the UHE band \citep{Amenomori2019, Abeysekara2020a}. Probing the TeV–PeV energy band is crucial for the source identification to explain the cosmic rays up to the knee and beyond.

The search for PeVatrons has been one of the key science topics for VERITAS and HAWC. VERITAS search has involved observations of young supernova remnants (SNR) such as Cas\,A \citep{Abeysekara2020b} and Tycho’s SNR \citep{Archambault2017}, in which spectral cut-offs were observed around $\sim$2\,TeV. In addition, observations of unidentified, hard-spectrum sources such as MGRO\,J1908+06, MGRO\,J2019+37, and VER\,J2227+608 (SNR\,G106.4+2.7 region) \citep{Aliu2014a, Aliu2014b, Acciari2009}, and HAWC sources from the second catalog \citep{Abeysekara2017} were also performed. In 2021, LHAASO detected 14 steady $\gamma$-ray sources with photon energies above 100\,TeV, up to 1.4\,PeV \citep{cao2021a, cao2021b, Aharonian2021}, probing the long-standing question whether hadronic PeVatrons exist. The LHAASO results provided a clear list of PeVatron candidates for further study with VERITAS and HAWC. Most of these sources contain possible source associations, such as SNRs, pulsar wind nebulae (PWN), or stellar clusters, except~LHAASO\,J0341+5258 \citep{cao2021b} and LHAASO J2108+5157 \citep{cao2021c}; which are without any such counterparts. Here we present a multi-wavelength study of LHAASO\,J0341+5258 to provide insight into the source nature and properties.

\section{Observations and Results} \label{sec:obs-ana}

In 2021, LHAASO reported the discovery of a new unidentiﬁed extended $\gamma$-ray source LHAASO\,J0341+5258 [R.A.\,=\,$(55.34\pm0.11)^{\circ}$ and DEC.\,=\,$(52.97\pm0.07)^{\circ}$] in the Galactic plane \citep{cao2021b} using 308.33 days of Kilometer Squared Array (KM2A) data. The source was detected with a significance of 6\,$\sigma$, an angular size of $(0.29\pm0.06_{stat}\pm0.02_{sys})^{\circ}$, and approximately 20\% of the Crab ﬂux above 25\,TeV. CO observations with the Milky Way Imaging Scroll Painting (MWISP) project \citep{Su2019} of the LHAASO J0341+5257 region show partially overlapping molecular gas in the form of a half-shell structure \citep{cao2021b}. The total mass of gas estimated using these data within 1$^{\circ}$ of the LHAASO source is about 10$^{3}$ M$_{\odot}$ assuming a distance of 1 kpc, and there is no clear CO emission observed at larger distances. The detection of a $^{13}CO$ line in the enhanced $^{12}CO$ emission region from the half-shell structure, even if the total CO emission is not bright in this region, indicates the existence of dense clumps.  

Recently, LHAASO published its first full catalog~\citep{cao2023} of 90 sources including 43 UHE sources above 100\,TeV using 508 days of data collected by the Water Cherenkov Detector Array (WCDA) and 933 days of data recorded by the KM2A. In this catalogue, they resolved LHAASO\,J0341+5258 into two sources using KM2A (1LHAASO\,J0339+5307 and 1LHAASO\,J0343+5254u*). Moreover, 1LHAASO\,J0343+5254u* was also detected in the 1-25\,TeV energy range using WCDA detector. Since it covers the same energy range as WCDA, VERITAS can, in principle, provide a complementary view of this source but with the added advantage of better angular and energy resolution.~1LHAASO\,J0343+5254u* was detected with a test statistic (TS) of 94.1 and had a similar extension to LHAASO\,J0341+5258 (0.33$\pm$0.05)$^{\circ}$. With KM2A, 1LHAASO\,J0339+5307 was detected at an offset from the position of LHAASO\,J0341+5258 of 0.37$^{\circ}$, with a TS=144, and extension$<$0.22$^{\circ}$, whereas 1LHAASO\,J0343+5254u* was detected at an offset of 0.28$^{\circ}$, TS=388.1 and extension of (0.20$\pm$0.02)$^{\circ}$.

\begin{figure*}
    %\centering % Not needed
    \begin{subfigure}[b]{0.44\columnwidth}
        \includegraphics[scale=0.47]{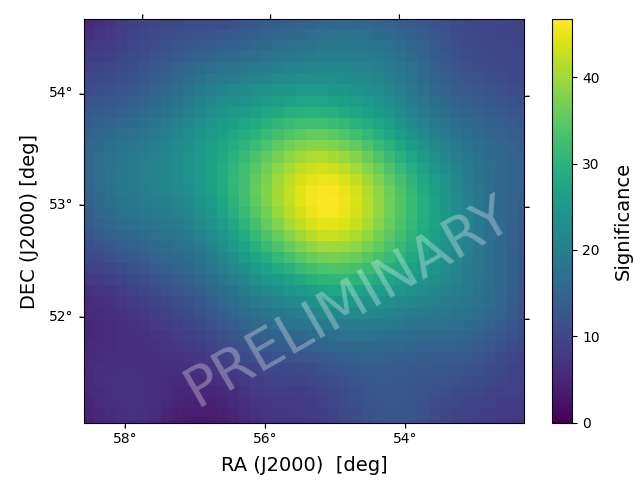}
        \caption{\textit{Fermi}-LAT}
        \label{fig:fermi_skymap}
    \end{subfigure}
    \hfill
    \begin{subfigure}[b]{0.49\columnwidth}
        \includegraphics[scale=0.46]{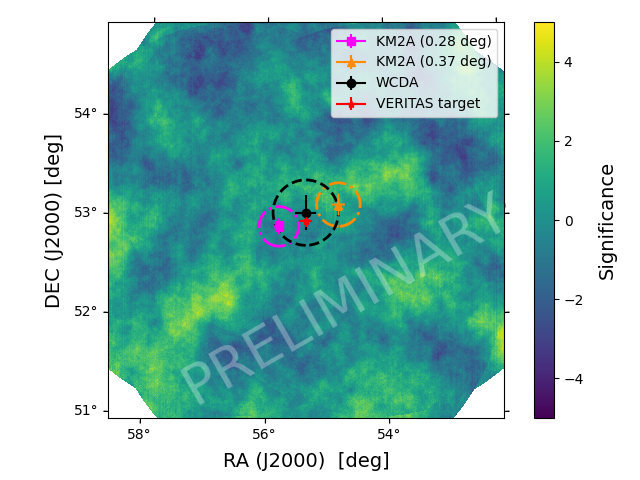}
        \caption{VERITAS}
        \label{fig:veritas_skymap}
    \end{subfigure}
    %% leave a blank line to create a line break
    \vfill
    \begin{subfigure}[b]{0.44\columnwidth}
        %\vspace{-0.5cm}
        \includegraphics[scale=0.44]{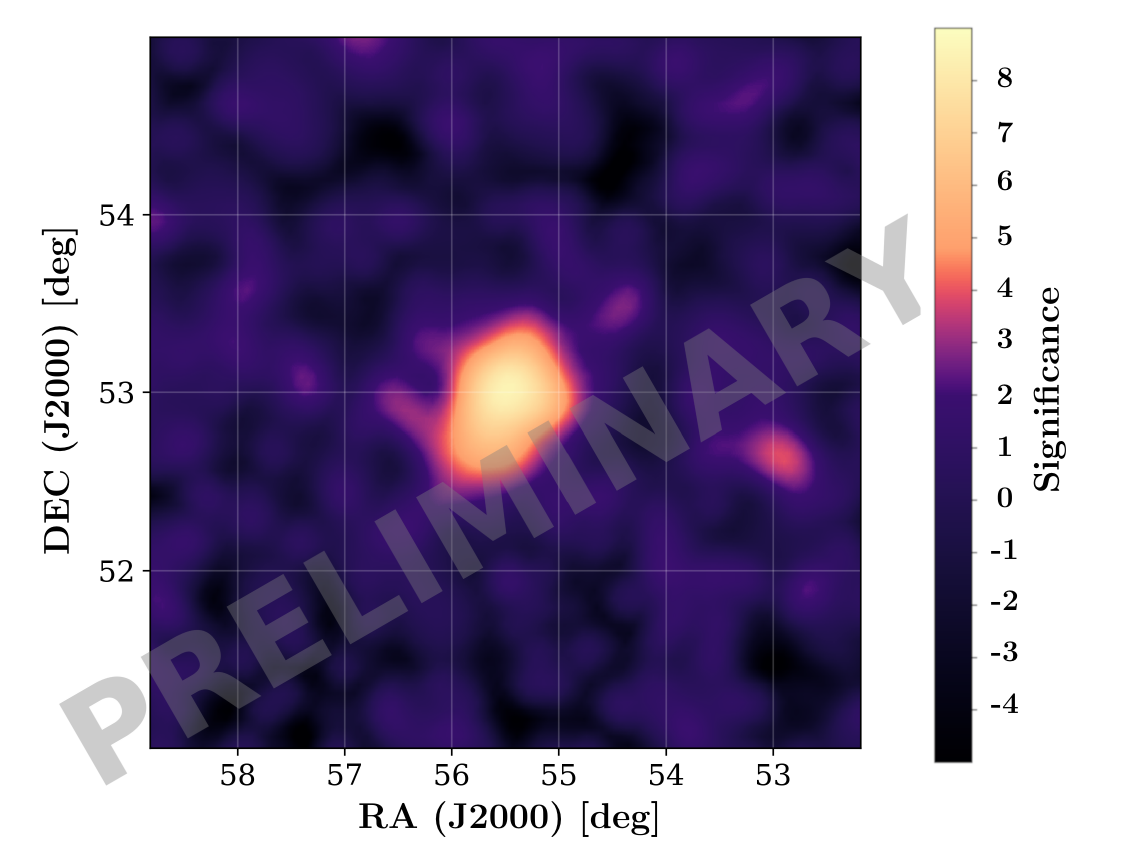}
        \caption{HAWC}
        
        \label{fig:hawc_skymap}
    \end{subfigure}
    \hfill
    \begin{subfigure}[b]{1\columnwidth}
    \end{subfigure}
    \caption{Significance maps for \textit{Fermi}-LAT, VERITAS, and HAWC data in the region centered on LHAASO\,J0341+5258 position. The source is detected clearly in \textit{Fermi}-LAT and HAWC data, but there is no significant detection in VERITAS data. The WCDA extension region is shown in a dashed black circle, and two extension regions corresponding to KM2A with the 0.28$^{\circ}$ and 0.37$^{\circ}$ offset are shown in dot-dashed magenta and orange circles, respectively, in the VERITAS significance map. Note that these extension regions are the 39\% containment radius of the 2D-Gaussian model (r\_39 region) reported in the LHAASO catalog \citep{cao2023}.}
    \label{fig:four figures}
\end{figure*}  

The closest unidentified GeV source, 4FGL J0340.4+5302 \citep{Abdollahi2017}, is located at an angular distance of 0.16$^{\circ}$ from the position of LHAASO\,J0341+5258 and is contained within the reported angular extension of LHAASO\,J0341+5258. Therefore, to investigate further, we have performed a dedicated binned likelihood analysis using fermipy v1.2 \citep{fermipy} for \textit{Fermi}-LAT pass8 data from Aug.\,2008 to Apr.\,2023. The region of interest was centered on the position of LHAASO\,J0341+5258 for $\gamma$ rays in the energy range of 100\,MeV to 1\,TeV. The events were binned into 0.1$^{\circ}$ spatial bins and two logarithmic energy bins per decade. The source is detected with a significance of $\sim$62\,$\sigma$ (TS=3888.3) (figure \ref{fig:fermi_skymap}). The best-fit central position is found to be 0.15$^{\circ}$ away from LHAASO J0341+5258, which is consistent with the position determined by LHAASO. The source is specified as point-like as the TS value of the extension hypothesis is $\sim$3.52, resulting in the 95\% upper limit of $\leq$0.20$^{\circ}$ on the extension. The spectrum of 4FGL J0340.4+5302 is significantly curved; therefore, we modeled it using a log-parabola function. The best-fit spectral parameters of the log-parabola model are an energy flux of (5.1\,$\pm$\,0.1)$\times \rm{10}^{-11}$ erg cm$^{-2}$ s$^{-1}$, spectral index $\alpha$\,=\,3.86\,$\pm$\,0.06, and $\beta$\,= 0.53\,$\pm$\,0.02 (figure \ref{fig:mwl_sed}). The overall spectrum is very soft, and no significant emission above 2 GeV is detected. 95\% C.L. upper limits are derived for the flux above 2 GeV.  

To investigate the TeV counterparts, we have searched VERITAS \citep{{Park2015}} and HAWC \citep{Abeysekara2017} data. VERITAS observations totalling 50 hrs were taken from Oct.\,2021\,-\,Jan.\,2022 and Sept.\,2022\,-\,Jan.\,2023, where observations were stopped in the gaps due to the source visibility and for the annual Arizona monsoon season.~The analysis was performed using the EventDisplay analysis package \citep{ED2017} and then cross-checked with the VEGAS analysis package \citep{vegas2008}. Considering the 0.33$^{\circ}$ source extension of LHAASO\,J0341+5258 from WCDA, we have searched VERITAS data with an extended source analysis using an integration region $\theta$ of 0.25$^{\circ}$ centered around the LHAASO\,J0341+5258 position. No significant emission was detected in these data (figure \ref{fig:veritas_skymap}). 99\% C.L upper limits were derived for three logarithmic energy bins per decade in the 0.5 to 31.6\,TeV energy range (figure \ref{fig:mwl_sed} and table \ref{table:veritas_ULs}).

\begin{table}[h!]
\centering
\begin{tabular}{||c| c||} 
 \hline
 Energy & VERITAS Upper limits       \\ [0.5ex] 
 [TeV] &   [TeV
 cm$^{-2}$ s$^{-1}$]   \\ [0.9ex]
 \hline\hline 
 1$_{-0.5}^{+2.0}$ & 2.79\,$\times$\,10$^{-13}$  \\ [1.2ex]
 \hline
 3.981$_{-2.0}^{+7.9}$ & 3.16\,$\times$\,10$^{-13}$  \\ [1.2ex]
 \hline
15.849$_{-7.9}^{+31.6}$  & 1.03\,$\times$\,10$^{-12}$  \\ [1.2ex]
  \hline
\end{tabular}
\caption{The 99\% C.L upper limits are derived from the 50 hrs of VERITAS data using an integration region of 0.25$^{\circ}$ centered around the LHAASO\,J0341+5258 position.}
\label{table:veritas_ULs}
\end{table}

For the HAWC analysis, we used 2582 days of fHit dataset from Nov.\,2014 to June\,2022. We binned the data using the fraction of available PMT channels\footnote{We calculate fraction of PMT as PMT\_triggered/PMT\_available for each event}, as described in the HAWC Crab analysis from 2017 \citep{Abeysekara2017}.~We choose a circular region of interest with a radius of 3$^{\circ}$ around the LHAASO\,J0341+5258 position. The source was detected with a significance of $\sim$8.4\,$\sigma$ (figure \ref{fig:hawc_skymap}). Both the point source model and extended source model with a simple power law are tested, but the extended source template is not preferred over point-source template. The 95\% C.L. upper limit on extension is 0.20$^{\circ}$. Hence further analysis was performed using a point source assumption. The resulting spectrum was fitted with a simple power-law using a spectral index of -2.24$\pm$0.22, normalization of 1.16$_{-0.24}^{+0.32}\,\times \rm{10}^{-13}$ TeV cm$^{-2}$ s$^{-1}$ for pivot energy of 40\,TeV. Further study of the source morphology and the shape of the energy spectrum is ongoing.

\section{Discussion}

\begin{figure}
    \centering
       \includegraphics[scale=0.5]{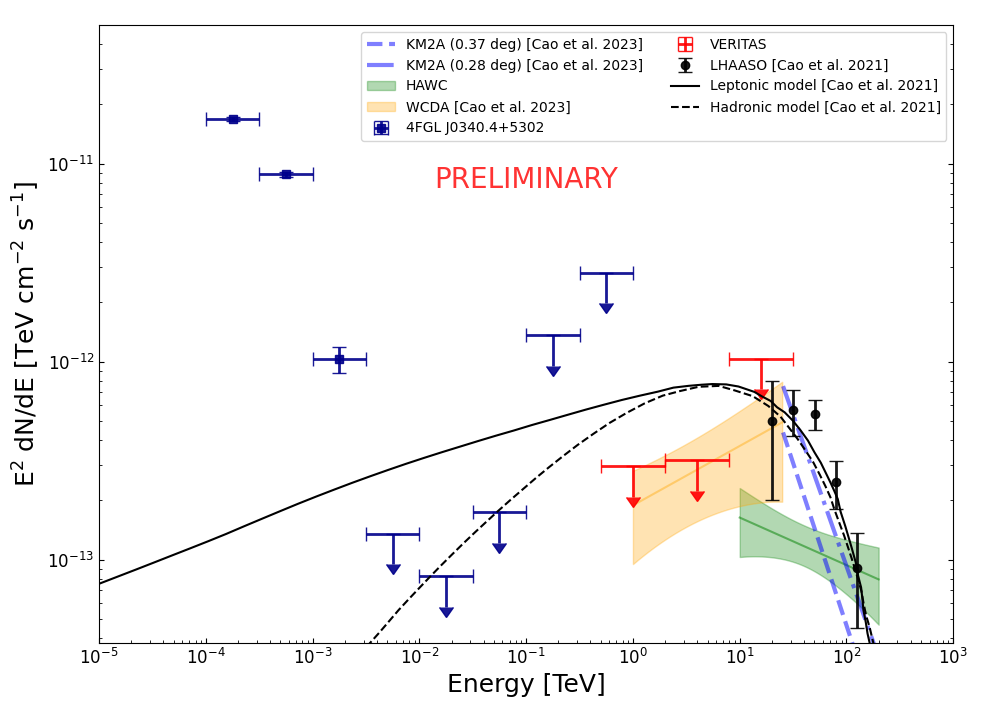}
        \caption{Multiwavelength spectral energy distribution using the \textit{Fermi}-LAT, VERITAS, HAWC, and LHAASO data.~VERITAS upper limits of 99\% C.L. (red) are derived using an integration region of 0.25$^{\circ}$ centered around the LHAASO\,J0341+5258 position. Note that  these limits are not directly comparable with the LHAASO flux points due to the differences in the source integration region (see text for more details).} 
        \label{fig:mwl_sed}
\end{figure}

The multi-wavelength spectral energy distribution (SED) is built using \textit{Fermi}-LAT, VERITAS, HAWC, and LHAASO data as shown in figure \ref{fig:mwl_sed}. We have also added WCDA and KM2A spectra provided by the first LHAASO catalog for comparison. Clearly, VERITAS upper limits already prove important, as they exclude the existing leptonic and hadronic emission models provided by LHAASO \citep{cao2021b}. In addition, these limits are compatible with the recently-reported WCDA spectral band, confirming the hard spectrum of LHAASO\,J0341+5258 when extrapolating toward low energies. One important subtlety to note is that the VERITAS upper limits are not directly comparable to LHAASO, as the VERITAS integration region is significantly smaller than the 39\% containment radius of the 2D-Gaussian model (r\_39) region reported by WCDA. Using a smooth, symmetrical 2D-Gaussian form for the emission, as assumed by WCDA, approximately $\sim$25\% of the total $\gamma$-ray flux is contained within the VERITAS integration region. However, in reality the situation is likely more complex, as the source region may include emission components with significant asymmetries, non-uniformities, and multiple source components.~The selection of integration region for VERITAS was based on the original LHAASO\,J0341+5258 results from \citep{cao2021b} and on considerations of signal-to-background ratio. For the current HAWC spectral analysis, using a simple power-law, the HAWC spectral band lies between the energy ranges of WCDA and KM2A. The LHAASO catalog paper indicates a possible transition of spectral indices in this region, around tens of TeV. HAWC observations and spectral fitting should allow to study and refine the measurement of this transition. In the future, we plan to use a log-parabola as a fitting function to test for possible spectral curvature, which might provide more consistency with LHAASO data.

Given the spatial coincidence between 4FGL J0340.4+5302 and LHAASO\,J0341+5258, one possible scenario is that both the 4FGL and LHAASO sources have a common origin. Moreover, the 4FGL J0340.4+5302 spectrum shows a sharp cutoff around 2 GeV, a typical signature of the GeV $\gamma$-ray emission of pulsars.~This implies that the TeV emission is more likely of leptonic origin, resulting from the inverse Compton scattering of relativistic electrons in the surrounding PWN or a pulsar halo \citep{Aharonian2004, Sudoh2019}. The caveat for this scenario is the lack of detection of a powerful pulsar, which could be due to the absence of a pulsar, misalignment of the pulsar beam to the line of sight of the observer, or the spin-down luminosity limit of pulsar being below the level required to produce detectable emission \citep{Guillemot2015}. As a molecular cloud partially overlaps LHAASO J0341+5257 region, an alternative possibility is a hadronic emission scenario, which could be explained by an old SNR, where cosmic rays have already escaped the SNR, encountering a nearby molecular cloud and being accumulated in a nearby but presently invisible SNR \citep{Gabici2007}. Our next step is to perform detailed multiwavelength spectral modeling to investigate the various possible emission scenarios, where the upper limits from VERITAS and \textit{Fermi}-LAT will play a key role.

\acknowledgments
This research is supported by grants from the U.S. Department of Energy Office of Science, the U.S. National Science Foundation and the Smithsonian Institution, by NSERC in Canada, and by the Helmholtz Association in Germany. This research used resources provided by the Open Science Grid, which is supported by the National Science Foundation and the U.S. Department of Energy's Office of Science, and resources of the National Energy Research Scientific Computing Center (NERSC), a U.S. Department of Energy Office of Science User Facility operated under Contract No. DE-AC02-05CH11231. We acknowledge the excellent work of the technical support staff at the Fred Lawrence Whipple Observatory and at the collaborating institutions in the construction and operation of the instrument. 

We acknowledge the support from: the US National Science Foundation (NSF); the US Department of Energy Office of High-Energy Physics; the Laboratory Directed Research and Development (LDRD) program of Los Alamos National Laboratory; Consejo Nacional de Ciencia y Tecnolog\'{i}a (CONACyT), M\'{e}xico, grants 271051, 232656, 260378, 179588, 254964, 258865, 243290, 132197, A1-S-46288, A1-S-22784, CF-2023-I-645, c\'{a}tedras 873, 1563, 341, 323, Red HAWC, M\'{e}xico; DGAPA-UNAM grants IG101323, IN111716-3, IN111419, IA102019, IN106521, IN110621, IN110521 , IN102223; VIEP-BUAP; PIFI 2012, 2013, PROFOCIE 2014, 2015; the University of Wisconsin Alumni Research Foundation; the Institute of Geophysics, Planetary Physics, and Signatures at Los Alamos National Laboratory; Polish Science Centre grant, DEC-2017/27/B/ST9/02272; Coordinaci\'{o}n de la Investigaci\'{o}n Cient\'{i}fica de la Universidad Michoacana; Royal Society - Newton Advanced Fellowship 180385; Generalitat Valenciana, grant CIDEGENT/2018/034; The Program Management Unit for Human Resources \& Institutional Development, Research and Innovation, NXPO (grant number B16F630069); Coordinaci\'{o}n General Acad\'{e}mica e Innovaci\'{o}n (CGAI-UdeG), PRODEP-SEP UDG-CA-499; Institute of Cosmic Ray Research (ICRR), University of Tokyo. H.F. acknowledges support by NASA under award number 80GSFC21M0002. We also acknowledge the significant contributions over many years of Stefan Westerhoff, Gaurang Yodh and Arnulfo Zepeda Dominguez, all deceased members of the HAWC collaboration. Thanks to Scott Delay, Luciano D\'{i}az and Eduardo Murrieta for technical support.

%---------------- Author list ----------------------------------

\clearpage

%\section*{Full Author List: VERITAS Collaboration}

\begin{center} 
\section*{\centering{All Authors and Affiliations}}

%---------------- veritas authors ----------------------------------
\normalsize{VERITAS COLLABORATION  \\}

\scriptsize
\noindent
A.~Acharyya$^{1}$,
C.~B.~Adams$^{2}$,
A.~Archer$^{3}$,
P.~Bangale$^{4}$,
J.~T.~Bartkoske$^{5}$,
P.~Batista$^{6}$,
W.~Benbow$^{7}$,
J.~L.~Christiansen$^{8}$,
A.~J.~Chromey$^{7}$,
A.~Duerr$^{5}$
% ...

\vspace{1cm}

%---------------- veritas affiliations ----------------------------------

\noindent
$^{1}${Department of Physics and Astronomy, University of Alabama, Tuscaloosa, AL 35487, USA}

$^{2}${Physics Department, Columbia University, New York, NY 10027, USA}

$^{3}${Department of Physics and Astronomy, DePauw University, Greencastle, IN 46135-0037, USA}

$^{4}${Department of Physics and Astronomy and the Bartol Research Institute, University of Delaware, Newark, DE 19716, USA}

$^{5}${Department of Physics and Astronomy, University of Utah, Salt Lake City, UT 84112, USA}

$^{6}${DESY, Platanenallee 6, 15738 Zeuthen, Germany}

$^{7}${Center for Astrophysics $|$ Harvard \& Smithsonian, Cambridge, MA 02138, USA}

$^{8}${Physics Department, California Polytechnic State University, San Luis Obispo, CA 94307, USA}

% ...

\end{center}

%-------------------------------------------------------

\end{document}